\input{aipcheck}

\documentclass[
final            
  ]
  {aipproc}

\layoutstyle{6x9}

\begin{document}

\title{Search for nuclearites with the ANTARES detector}

\classification{21.65.Qr,14.80.-j,95.55Vj,96.50.sf}
\keywords      {quark matter, nuclearites, neutrino telescopes}

\author{G.E. P{\v a}v{\v a}la{\c s}}{
  address={Institute for Space Sciences, Bucharest-Magurele,Romania}
}


\begin{abstract}
	ANTARES is an underwater detector located in the Mediterranean Sea, near the French city of Toulon, dedicated to the search for cosmic neutrinos. ANTARES is optimized to detect the Cherenkov signal from up-going relativistic particles, but could also observe massive exotic objects, such as magnetic monopoles and nuclearites. In this article we present a search strategy for nuclearites and determine the sensitivity to nuclearites of ANTARES detector in complete configuration, using a set of data taken in 2008. 

\end{abstract}

\maketitle


\section{Introduction}

Strange quark matter, composed of nearly equal number of up, down and strange quarks, could be the ground state of nuclear matter \cite{Witten}.
 Massive lumps of strange quark matter, named "nuclearites" by Glashow and De Rujula (1984), could be present in cosmic radiation and reach the Earth \cite{ruj}. Nuclearites may be produced in the Early Universe or as debris from supernovae and strange stars collisions \cite{Madsen}. They could be detected while traveling through transparent media (water, air), where they transfer some of their energy as visible light. 

 ANTARES is a neutrino telescope aimed to detect high energy neutrinos from galactic and extragalactic sources, such as supernovae, binary systems and gamma ray bursts. The detector is optimized to collect the Cerenkov light emitted by up-going relativistic muons, produced in neutrino interactions below the detector. 
ANTARES could also be sensitive to the signal of down-going nuclearites \citep{Popa:2006,Pav:2007,2009arXiv0908.0860P}. 

The goal of this study is to develop a search strategy for nuclearites with the fully installed ANTARES telescope.

\section{The ANTARES detector}

\paragraph{General description}
	ANTARES is located on the floor of the Mediterranean Sea, at a depth of 2.5 km. It consists of a three dimensional array of 885 photomultiplier tubes (PMTs) distributed on 12 lines anchored on the seabed. The detector is operated from a control room on shore. The sensitive element of the ANTARES telescope is a hemispherical 10'' Hamamatsu photomultiplier tube,  housed in a pressure resistant glass sphere \cite{Aguilar:2005}. 
A triplet of PMTs forms a storey, together with the read-out and control electronics. A detector line has a length of 450 m and contains 25 storeys, distributed at every 14.5 m, starting 100 m above the seabed. Every line is connected to the Junction Box, itself connected to the shore station at La Seyne-sur-Mer through a 40 km long electro-optical cable. The detector instruments a surface area of about 0.1 km$^{2}$. 

\paragraph{Data acquisition system}

The PMT signals larger than a preset threshold of 0.3 photoelectrons (pe) are digitized and the corresponding time and charge informations, referred to as L0 hits, are sent to shore. A L1 hit is defined either as a local coincidence of L0 hits on the same storey within 20 ns, or as a single hit with a large amplitude, typically 3 pe. The raw data are processed by a computer farm in the shore station, using different triggers for physics signals \cite{Aguilar:2007}. 

Two muon triggers operated during data acquisition in 2008: the {\it directional} trigger and the {\it cluster} trigger.
The directional trigger requires five local coincidences causally connected, within a time window of 2.2 $\mu$s, and the cluster trigger requires two coincidences between two L1 hits in adjacent or next-to-adjacent storeys.  When a muon event is triggered, all PMT pulses are recorded over 4 $\mu$s in a {\it snapshot}.\\

ANTARES was taking data in partial configurations since March 2006 and was completed in May 2008.
Measurements of the atmospheric muon flux with the first ANTARES line and the 5-line detector are presented in \cite{2009APh....31..277A}, and respectively \cite{Aguilar:2010}.

\section{Nuclearites}

Nuclearites are hypothetical massive particles, composed of nearly equal numbers of up, down and strange quarks \cite{ruj}. They are believed to be neutralized by electrons inside the quark core or by an external electron cloud, forming a sort of an atom.
Their velocity is non-relativistic  ($\beta=10^{-3}$) and interact with the surrounding medium via elastic and quasi elastic collisions, with an energy loss:  $dE/dx = -\sigma\rho{v^2}$,
where $\rho$ is the density of the medium, $v $ is the nuclearite velocity and $\sigma$ its geometrical cross section:
 
   \begin{displaymath}
        \sigma = \left\{ \begin{array}{ll}
                        \pi(3M_N/4\pi\rho_N)^{2/3} & \mbox{for $M_N\geq8.4*10^{14}$ GeV};\\
                       \pi\times10^{-16} \mbox{cm}^2 &  \mbox{for lower masses},
                    
	             \end{array}
	             \right. 
	\label{sigma_cross}
     \end{displaymath}
where $\rho_N=3.6\times10^{14}$ g cm$^{-3}$ is the density of a nuclearite.
Nuclearites moving slowly through water would produce a thermal shock emitting blackbody radiation. The luminous efficiency was estimated to be $\eta\simeq3\times10^{-5}$ by \cite{ruj} and the number of visible photons emitted per unit path length can be computed from: ${dN_{\gamma}}/{dx}=\eta\frac{dE/dx}{\pi(eV)}$,
assuming a mean energy of visible photons of $\pi$ eV.   

\section{Data and Monte Carlo simulations}

\paragraph{Monte Carlo simulations}
The Monte Carlo simulations of nuclearite events in ANTARES use a hemispherical generation volume of 548 m radius, that surrounds symmetrically the detector \cite{Popa:2005}. The initial point of the trajectory and the direction of the nuclearite are randomly generated.The algorithm proceeds in steps of 2 ns, evaluating the position and $\beta$ of nuclearites, as well as the number of hits on the PMTs. 
We have simulated a sample of about 350000 down-going nuclearites in 12 line configuration, with an initial velocity (before entering the atmosphere) of $\beta=10^{-3}$, in the mass range from $10^{15}$ to $10^{18}$ GeV . 
A typical nuclearite would cross the ANTARES detector in a characteristic time of ~1 ms, producing a large light signal.

The main background for nuclearites is represented by the down-going atmospheric muons. The atmospheric muon sample used in this analysis was generated with the MUPAGE code [9], considering an energy range from 20 GeV to 500 TeV.  Other sources of background are the decays of K40 (with a constant rate of about 27 kHz), and the bioluminiscence (characterized by large fluctuations in time).

The MC samples are processed with the same triggers used in data acquisition, and the background is added from selected runs. Processing the nuclearite events with triggers designed for relativistic particles results in a sequence of snapshots, distributed on a detector crossing timescale up to  three orders of magnitude larger than for muons. 

\paragraph{Data}
A period of 42.8 days of data taken in 12 line configuration, from June to December 2008, is considered in the analysis.
Both the directional and the cluster triggers were operated during the data acquisition, with a high threshold of 3 pe. The selected data satisfies certain quality criteria, such as a low biological activity.

\section{Analysis and results}
	
\paragraph{Analysis method} 
The analysis uses a {\it blinding} policy, that proceeds according to the following steps: 
\begin{enumerate}
\item look for the best discriminative variables, then define selection cuts for nuclearites. At this step we use simulated nuclearites and muons;
\item comparison between a sample of 15\% of data and Monte Carlo down-going muons, in order to validate the simulations; 
\item calculation of the ANTARES sensitivity to nuclearites for 42.8 days of 12 line data.
\end{enumerate}
      
\paragraph{Definition of the selection conditions}           
In the first step, we studied the distributions of several parameters for the Monte Carlo nuclearite and atmospheric muon samples, such as the number of L0 hits and L1 hits per snapshot and the duration of the cluster of L1 hits, defined as the time difference between the  last and the first L1 hit in the snapshot, see Figs. 1-2.  The distributions are normalized to unity. 
          
The distributions in Fig. 1 are showing an excess of L0 and L1 signals per snapshot for nuclearites (represented with dash-dotted line) with respect to the simulated atmospheric muons (dotted line). This fact allowed us to define a first selection criterion for nuclearites as a linear combination the number of L0 hits and L1 hits. This cut is shown in Fig. 2 (right), that separates the nuclearite snapshots (represented as points) from the simulated muons (crosses). A small percentage of the actual simulated events is used, for the legibility of the figure. 
\begin{figure}
  \includegraphics[height=.22\textheight]{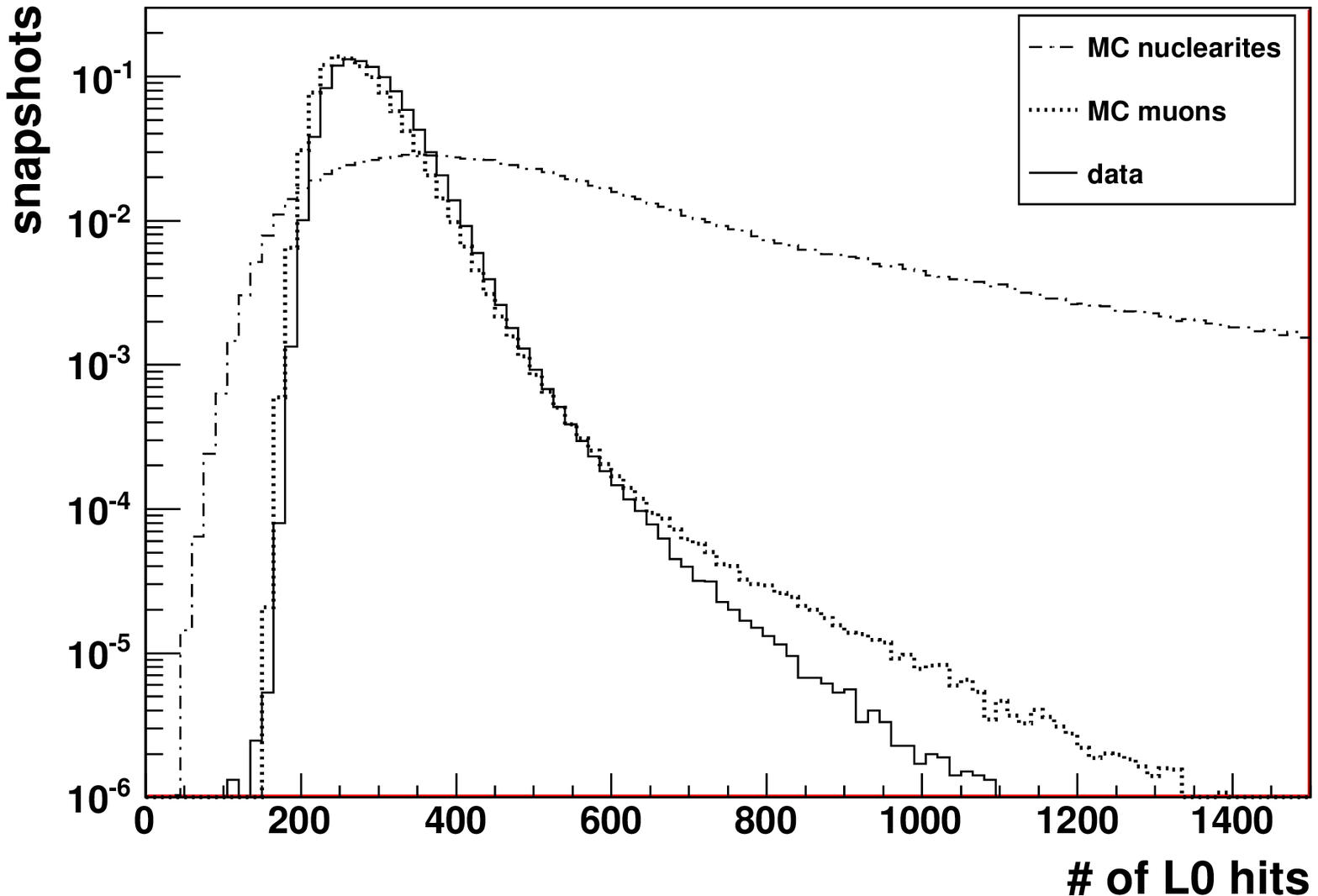}
  \includegraphics[height=.22\textheight]{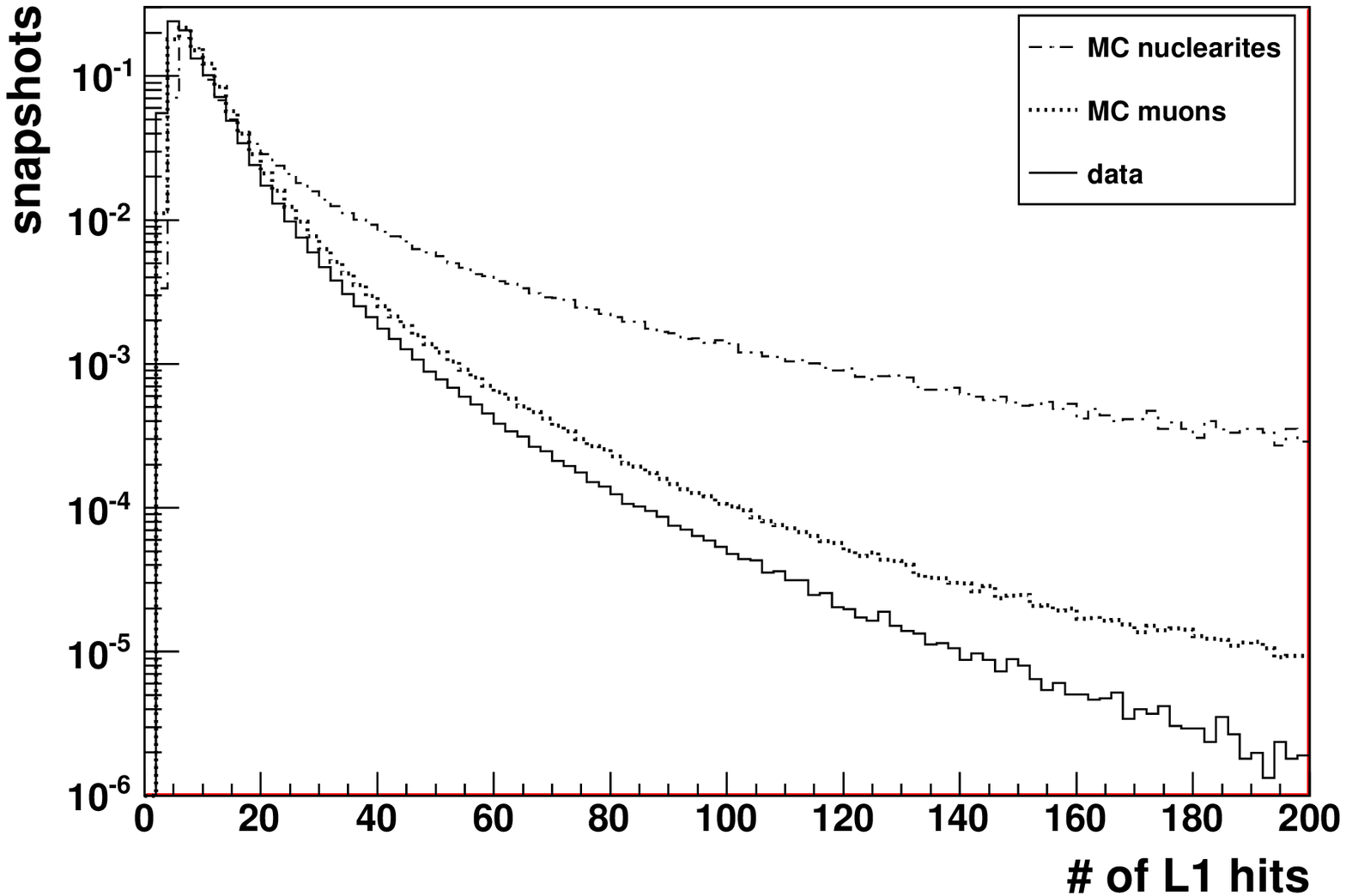}

    \caption{Normalized distributions of the number of L0 hits (left) and L1 hits (right) per snapshot for simulated nuclearites (dash-dot line), muons (dotted line) and 15\% of data (continuous line).}
\end{figure}	

\begin{figure}
  \includegraphics[height=.22\textheight]{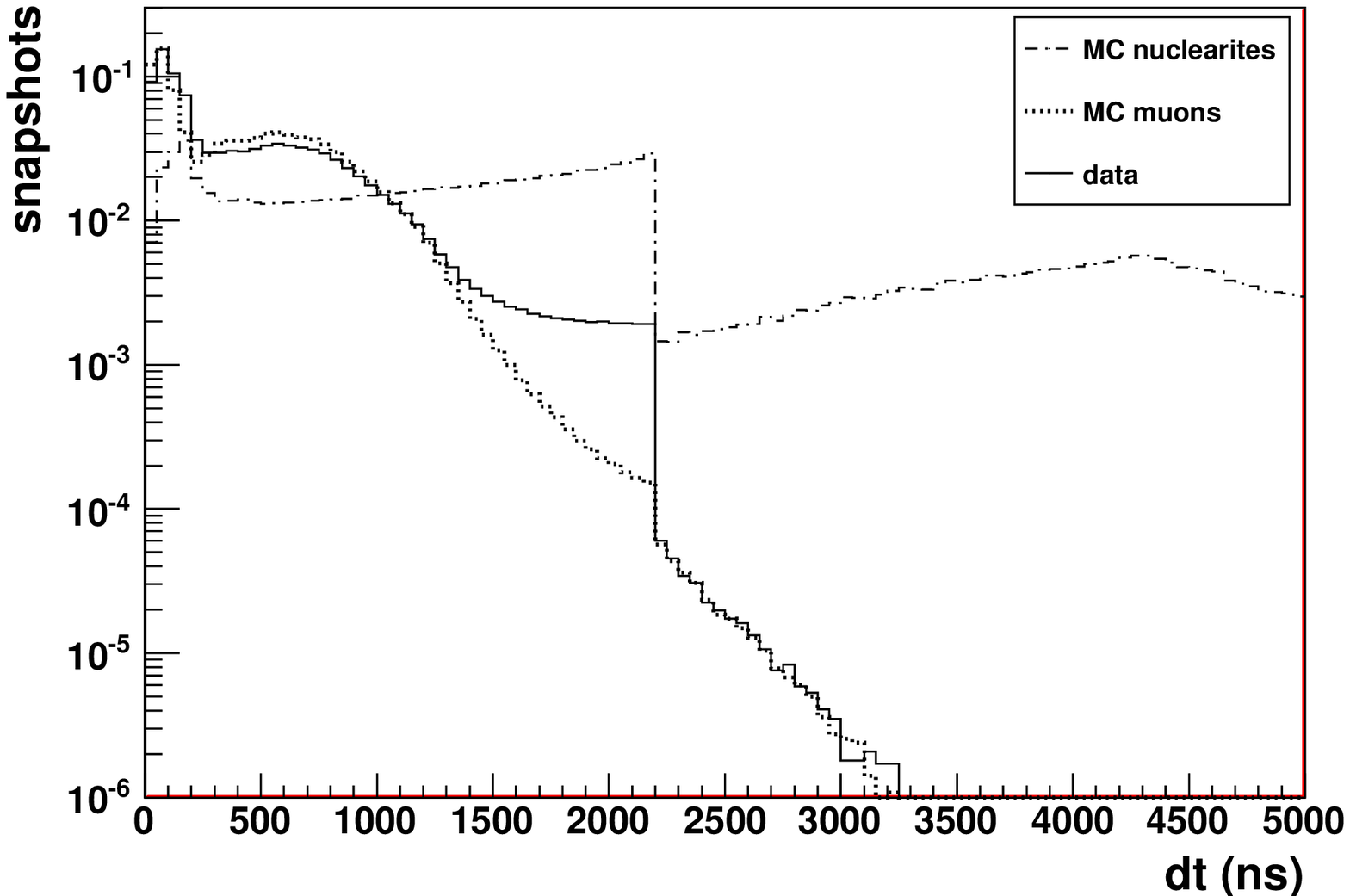}
    \includegraphics[height=.22\textheight]{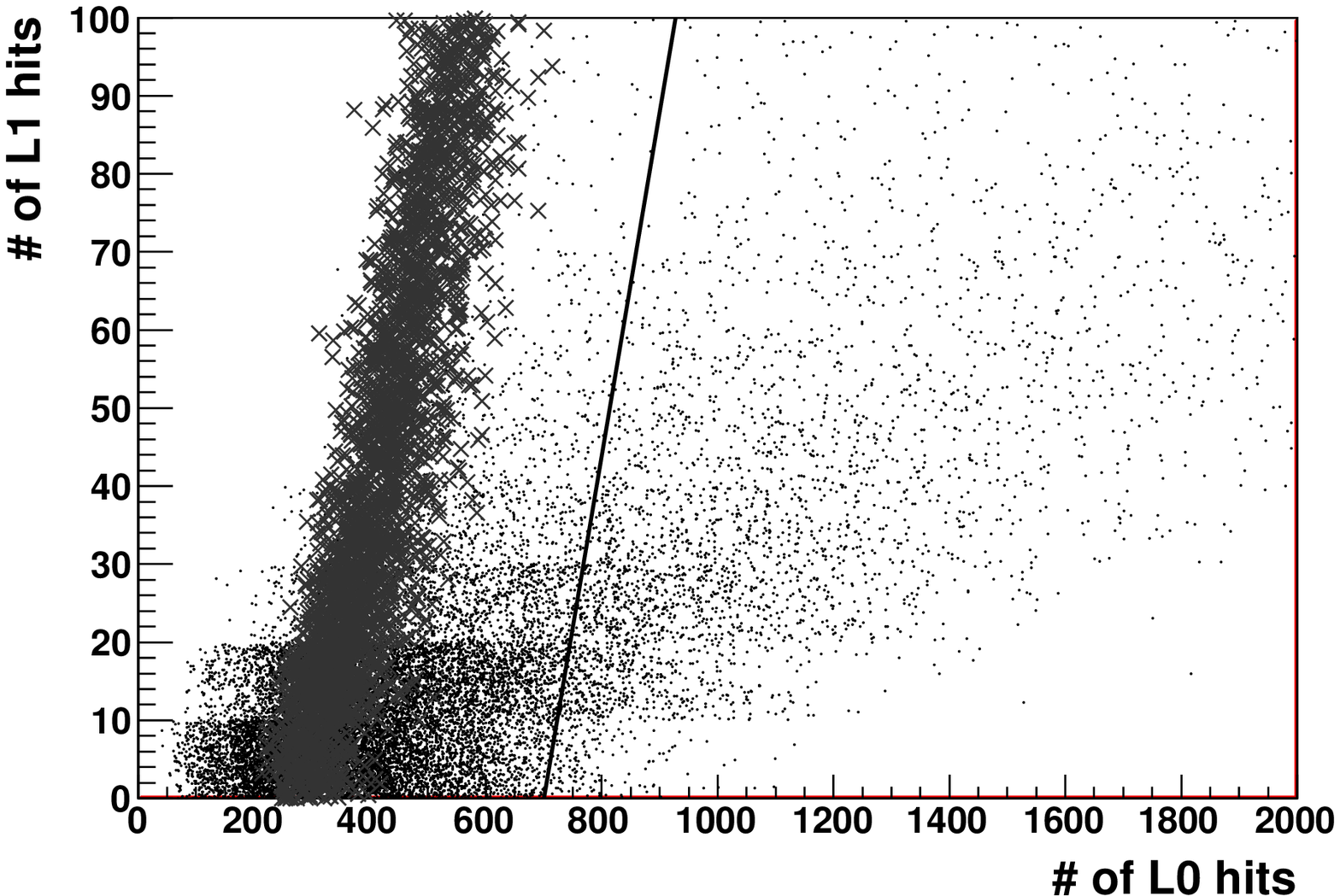}
  \caption{Left: Normalized distributions of the L1 cluster duration (dt) per snapshot for simulated nuclearites and atmospheric muons. The distribution from 15\% of data is also shown. Right: Linear selection cut in the distribution of the number of L1 hits as  a function of L0 hits for nuclearite snapshots (points) and atmospheric muons (crosses).}
\end{figure}
          
\paragraph{Data - Monte Carlo comparison}        
The comparison between data and Monte Carlo atmospheric muons is also shown in Fig. 1 and Fig. 2 (left), using 15\% of the data taken with 12 lines in 2008, that is equivalent to 6.4 days of active time. The data and Monte Carlo distributions are in relatively good agreement, with some discrepancies in the tails of the distributions.

	The efficiency of the first cut on simulated nuclearites considering all masses is relatively high, about 74\%, while the rejection efficiency for simulated down-going muons is 100\%. After applying this cut on the data sample, two events remain. 
	
	In order to optimize the rejection of the two remaining events in data, a second cut is applied, that requires two or more snapshots within 1 millisecond, with the duration of the L1 hits cluster larger than 3000 ns. 
	The rejection efficiency of the two combined cuts is 100\% on the 15\% data sample, while the overall efficiency for nuclearites, computed with all masses, drops to 73\%.
	Nuclearite events with $10^{15}$ GeV mass do not pass the selection cuts.
	
\paragraph{Expected sensitivity}	
In the final step, we computed the sensitivity to nuclearites. The sensitivity is defined as the 90\% C.L. flux upper limit that we expect for a given background prediction, and no true signal \cite{Feld}. The obtained sensitivity to down-going nuclearites for a period of 42.8 days of 12 line ANTARES data and no background events varies between $4.1\times 10^{-16}$ cm$^{-2}$ s$^{-1}$ sr$^{-1}$  for $10^{16}$ GeV nuclearites and $5.5\times 10^{-17}$ cm$^{-2}$ s$^{-1}$ sr$^{-1}$  for $10^{18}$ GeV.

The sensitivity obtained for down-going nuclearites of lower masses is comparable with the best upper flux limit established by the MACRO experiment, of $2.7\times10^{-16}$ cm$^{-2}$ s$^{-1}$ sr$^{-1}$ \cite{MACRO}.  The result obtained for $10^{18}$ GeV nuclearite mass improves the MACRO result by about a factor 5.

\section{Conclusions}
ANTARES has the instrumental capability to search for massive exotic particles. In this study, we have determined the sensitivity of ANTARES to nuclearites. These preliminary results show that the sensitivity at lower masses is comparable with the MACRO upper limit, and is improved by a factor 5 at large masses.
ANTARES sensitivity to nuclearites will improve significantly with more data.


\begin{theacknowledgments}
The author would like to thank to  Prof. Giorgio Giacomelli and Dr. Vlad Popa for useful discussions and support.
 \end{theacknowledgments}



\bibliographystyle{aipproc}   
\bibliography{sample}

\end{document}